\documentstyle[epsfig,epsf]{article}
\def\nmne {\nu_\mu \longleftrightarrow \nu_e}
\def\nmnt {\nu_\mu \longleftrightarrow \nu_\tau}

\def\ne {\nu_e}
\def\nm {\nu_\mu}
\def\nt {\nu_\tau}
\def\dm2 {\Delta m^2}
\def\s2t {\sin^2 2\theta}

\begin{document}

\begin{center}
{\Large {\bf THE OPERA EXPERIMENT}}
\end{center}

\vskip .7 cm

\begin{center}
G. GIACOMELLI and M. GIORGINI \par
for the OPERA Collaboration \par~\par
{\it Dept of Physics, Univ. of Bologna and INFN, \\
V.le C. Berti Pichat 6/2, Bologna, I-40127, Italy\\} 

E-mail: giacomelli@bo.infn.it , giorginim@bo.infn.it

\par~\par

Talk given at Vulcano Workshop 2006, Frontier Objects in Astrophysics and  
Particle Physics, Vulcano, Italy, May 2006.

\vskip .7 cm
{\large \bf Abstract}\par
\end{center}

{\normalsize 
The physics motivations and the detector design of 
the long baseline OPERA experiment are discussed; OPERA is a hybrid
detector made of several types of electronic subdetectors, 2 magnets and
lead/nuclear emulsions ``brick'' walls. It is located in the Gran Sasso 
underground lab, 732 km from CERN, on the CNGS neutrino beam. A summary 
of the performances and of the physics plans are presented.}

\vspace{5mm}

\large
\section{Introduction}\label{sec:intro}
Neutrino physics has opened new windows into phenomena beyond the Standard 
Model of particle physics. Long baseline neutrino experiments may allow
further insight into neutrino physics. The CERN to Gran Sasso neutrino beam 
(CNGS) is one of these projects \cite{CNGS}, and one of the main experiments 
is OPERA \cite{opera}, designed to search 
for the $\nmnt$ oscillations in the 
parameter region indicated by the MACRO \cite{macrobib}, SuperKamiokande 
\cite{skbib} and Soudan2 \cite{soudanbib} 
atmospheric neutrino results \cite{gg-io}, recently confirmed
by the K2K \cite{k2k} and MINOS \cite{minos} experiments.
 The main goal of OPERA is to find the $\nt$ appearance by direct 
detection of the $\tau$ lepton from $\nt$ CC interactions. One may 
also search for the subleading $\nmne$ oscillations and make a 
variety of observations with or without the beam using the electronic 
subdetectors. 

\begin{figure}[h!]
\begin{center}
\mbox{\epsfig{figure=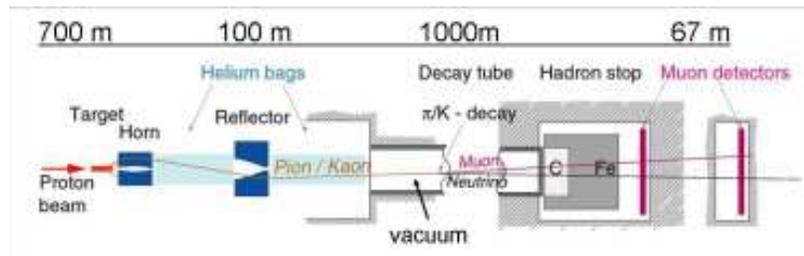,height=3.4cm}}
\caption {The main components of the CNGS neutrino beam.}
\label{fig:CNGS}
\end{center}
\end{figure}

Assuming a beam intensity of $4.5\cdot 10^{19}$ pot/year 
and a five year run, 31000 neutrino events (CC + NC) are expected in 
an average target mass of 1.6 kt. 95 (214) $\nt$ CC 
interactions are expected for $\dm2 = 2 \cdot 10^{-3}$
eV$^2$ (for $\dm2 = 3 \cdot 10^{-3}$ eV$^2$). The detection of the 
$\nt$'s will be made via the charged 
$\tau$ lepton produced in $\nt$ CC interactions, and its decay products. The 
$\tau$ decays can be classified as muonic (BR 18\%), electronic (BR 18\%) 
and hadronic (BR 64\%). To observe the decays, a spatial resolution of 
$\sim 1~\mu$m is necessary; this resolution is obtained in emulsion 
sheets interspersed with thin lead target plates. This technique, the 
Emulsion Cloud Chamber (ECC), was started in the $\tau$ search 
experiment \cite{donutbib}. 

The basic target module is a {\it ``brick''}, consisting 
of a sequence of 56 lead plates (1 mm thick) and 57 emulsion 
layers. A brick has a size of $10.2 \times 12.7$ cm$^2$, a 
depth of 7.5 cm (10 radiation lengths) and a weight of 8.3 kg. Two 
additional emulsion sheets, the {\it changeable 
sheets}, are glued on its downstream face. The bricks are arranged in 
walls. Within a brick, the achieved spatial resolution is $< 1~\mu$m and 
the angular resolution is 2 mrad. These values allow the reconstruction of 
the $\nu$ interaction vertex and of the $\tau$ decay topology. 
 To provide a $\nu$ interaction trigger and to identify the brick in which 
the interaction took place, the brick walls are complemented by a target 
tracker and a muon spectrometer. The target tracker consists of 
highly segmented scintillator planes inserted between the brick 
walls; the magnetic spectrometer measures the muon 
momentum and identifies the sign of the charges. Combining the signals left 
in the electronic detectors, the 
brick containing the $\nu$ interaction vertex can be determined with a 
global efficiency of $\sim 9\%$. The brick is extracted  
and its changeable sheets are developed and scanned using fast automatic 
microscopes. If the presence of tracks from an interaction is 
confirmed, all the emulsion sheets of the brick are developed 
and sent to the scanning labs for further analysis.

\begin{figure}[h!]
\begin{center}
\mbox{\epsfig{figure=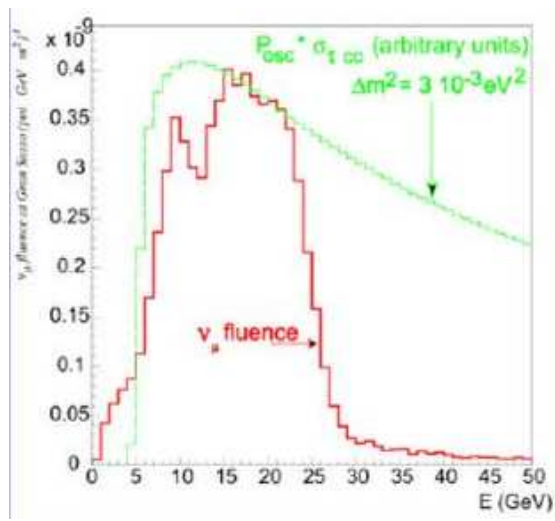,height=6.8cm}}
\caption {Spectrum of the CNGS $\nm$ beam at Gran Sasso.}
\label{fig:spettro}
\end{center}
\end{figure}

\section{The CNGS neutrino beam}
\label{sec:CNGS}
Fig. \ref{fig:CNGS} shows the main components of the $\nm$ beam 
at CERN \cite{CNGS}. The CNGS beam is optimised for a maximum number of CC 
$\nt$ interactions at Gran Sasso (732 km from CERN). The energy distribution
at Gran Sasso is shown in Fig. \ref{fig:spettro}: the primary $p$ energy is 
400 GeV, the mean $\nm$ beam energy is 17 GeV, the $\bar{\nm}$ 
contamination is $\sim 2\%$, the $\ne$ ($\bar{\ne}$) is $< 1\%$ and the 
number of $\nt$ is negligible. The $L/E_\nu$ ratio is 43 km/GeV. Civil 
engineering is completed, all beam parts are installed and commissioning 
is being made; the first low intensity beam is expected at GS in August 2006.

\section{The OPERA detector}
\label{sec:construction}
\begin{figure}
\begin{center}
\mbox{\epsfig{figure=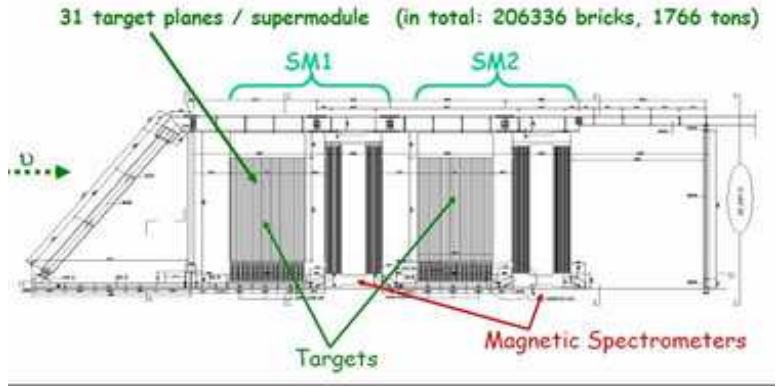,height=5.2cm}}
\caption {Layout of the OPERA detector.}
\label{fig:detopera}
\end{center}
\end{figure}

The OPERA detector, Fig. \ref{fig:detopera}, is made of two 
identical super-modules, each consisting of a target section 
with 31 target planes followed by a muon spectrometer. With  
206000 bricks, the initial target mass is 1.8 kt. 

{\bf Electronic subdetectors.} The first subdetector is an 
{\it anticoincidence wall} to better separate muon events coming from 
interactions in OPERA and in the material before.

The {\it target tracker} is made of 32000 scintillator strips, each 7 m 
long and of 25 mm$ \times $15 mm cross section (7000 m$^2$ area). Along 
the strip, a wavelength shifting 
fibre of 1 mm diameter transmits the light signals to both ends. The readout 
is done by 1000 64 channel HAMAMATSU PMTs. The target of the first 
super-module was installed in November 2005, the second in June 2006. A 
brick wall position accuracy is better than 1 mm. 

The {\it muon spectrometer} consists of 2 iron magnets instrumented with 
{\it Resistive Plate Chambers} (RPC) and {\it drift tubes}. Each magnet is an 
$8 \times 8$ m$^2$ dipole with a field 
of 1.55 T in the upward direction on one side and in the downward direction on 
the other side. This allows to measure the momentum twice, 
 reducing the error by $\sqrt{2}$. A magnet consists of 
twelve 5 cm thick iron slabs, alternated with RPC planes. In the magnetic 
field a muon is tracked, identified and its momentum is measured. 

The {\it precision tracker} 
\cite{specbib} measures the muon track coordinates in the horizontal 
plane. It is made of 12 drift tube planes, each covering an area of 
$8 \times 8$ m$^2$; they are placed in front and behind each 
magnet and between the two magnets. Each drift tube is 8 m long and has 
an outer diameter of 38 mm. 
 The efficiency of the muon identification, the accuracy of the 
momentum measurement and sign determination are increased; the charge 
misidentification should be $ 0.1 \div 0.3 \%$. This minimises the 
background from charmed particles produced in $\nm$ 
interactions. The muon spectrometer allows a momentum resolution 
$\Delta p / p \le 0.25$ for muon momenta $< 25$ GeV/c. 
 To reduce the number of ``ghost tracks'' two planes of {\it glass 
RPC's (XPC's}), consisting of two $45^{\circ}$ crossed planes, 
 are installed in front of the magnets.

\begin{figure}[h!]
\begin{center}
\mbox{\epsfig{figure=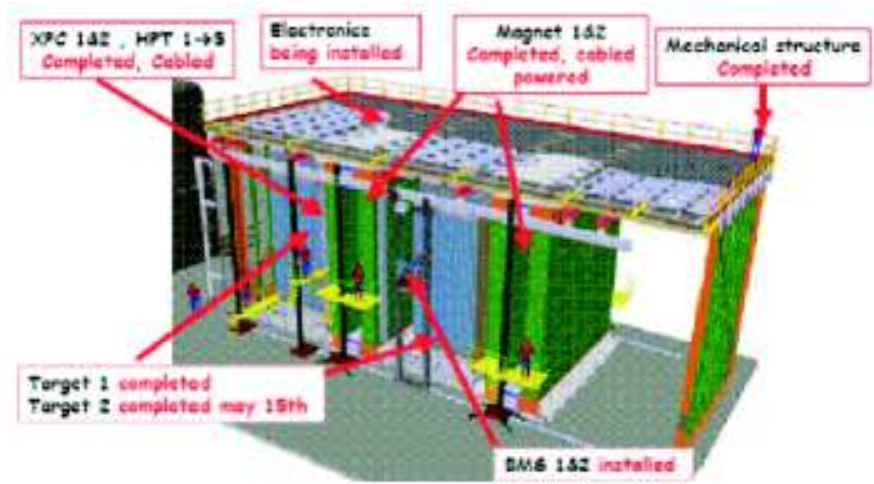,height=6.5cm}}
\caption {Status of detector installation (July 2006).}
\label{fig:status}
\end{center}
\end{figure}

The construction status in July 2006 is shown in Fig. 
\ref{fig:status}. The brick supporting structure, the tracker planes, 
 the XPC's and three of the high
precision tracker planes of the first supermodule are installed. The 
magnets, including all RPC's and the mechanical structure are completed. 

To handle the data flow a new DAQ system was developed. It uses a 
Gigabit network consisting of 1200 nodes. To match the data of the different 
subdetectors an event time stamp is delivered by a clock using the Global 
Positioning System (GPS). The DAQ uses a system which contains 
the CPU, the memory, the clock receiver for the time stamp and the ethernet 
connections to the other components. The components of the DAQ system 
are under test.

The commissioning of each subdetector is underway. The final commissioning 
 will be made with the CNGS at reduced intensity in August 2006.

{\bf Nuclear emulsions and their scanning.} The production of the bricks 
is made by the {\it Brick Assembling Machine} 
(BAM). It consists of robots for the mechanical packing of the bricks. In 
total 23 million lead and emulsion layers are needed to make the 
bricks. The final system is now installed in the Gran Sasso lab  
and its production speed is $\sim 2$ bricks per minute. 

The bricks are handled by the {\it Brick 
Manipulator System} (BMS), made of two robots, each operating at one 
side of the detector; one robot consists of a drum for brick transfer and a 
brick storage carousel. An arm is used to insert the bricks. The extraction 
of a brick, in the region  
indicated by the electronic detectors, is done by a vacuum sucker. The 
first BMS robot was installed in 2005 and the whole system is now being 
commissioned. 

A fast automated scanning system is needed to cope with the daily analysis 
of a large number of emulsion sheets. The minimum required scanning speed is 
$\sim 20$ cm$^2$/h per emulsion layer ($44 ~\mu$m thick). 
It corresponds to an increase in speed of at least one order of magnitude 
with respect to past systems \cite{TS,SYSAL}. For this purpose OPERA 
developed the {\it European Scanning System} (ESS) \cite{ESS} and the 
{\it S-UTS} in Japan \cite{SUTS}. 

\begin{figure}
\begin{center}
\mbox{\epsfig{figure=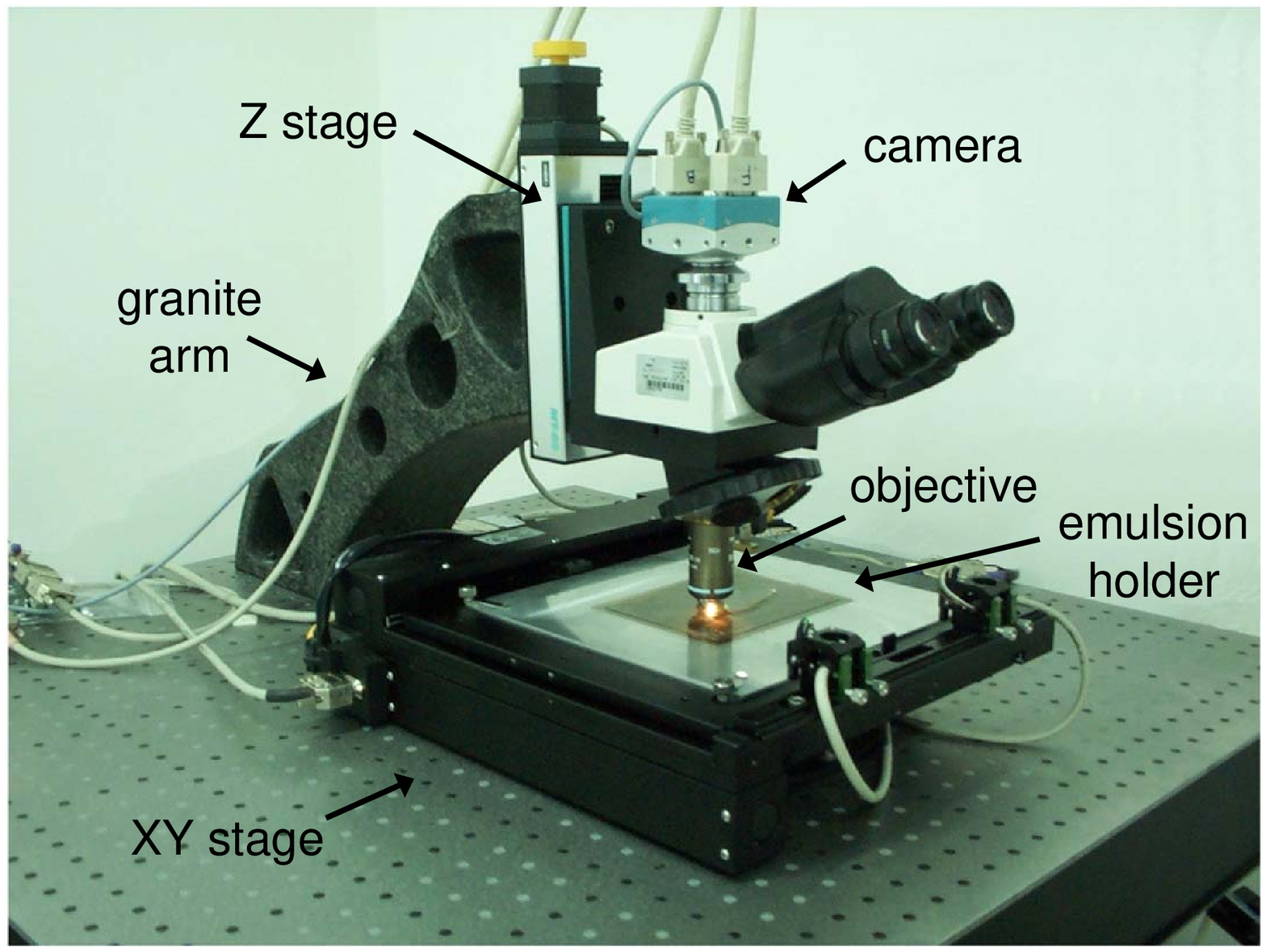,height=5.0cm}
      \epsfig{figure=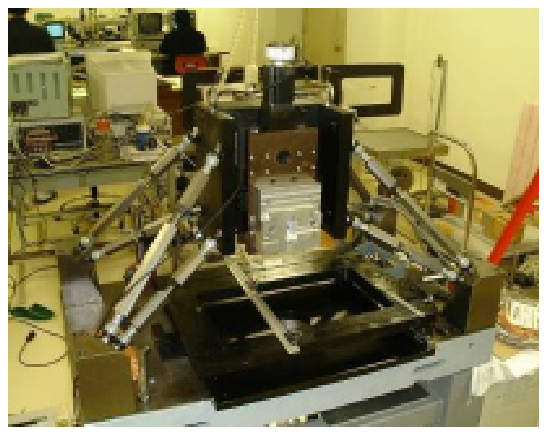,height=5.0cm} }
\caption {Photograph of one of the ESS microscopes (left) and of the
  S-UTS (right).}
\label{fig:ESS}
\end{center}
\end{figure}

The main components of the ESS microscope are shown in Fig. 
\ref{fig:ESS} left: 
 (i) a high quality, rigid and vibration-free support table; (ii) a motor 
driven scanning stage for horizontal (XY) motion; (iii) a granite arm; (iv) 
a motor driven stage mounted vertically (Z) on the granite 
arm for focusing; (v) optics; (vi) digital camera for image grabbing 
mounted on the vertical stage and connected with a vision processor; (vii) 
an illumination system located below the scanning table. The emulsion 
sheet is placed on a glass plate (emulsion holder) and its flatness is 
guaranteed by a vacuum system. 

By adjusting the focal plane of the objective, the $44~\mu$m 
emulsion thickness is spanned and 16 tomographic images 
of each field of view, taken at equally spaced depth levels, are 
obtained. The images are digitized, converted into a grey scale 
of 256 levels, sent to a vision processor board and analyzed to recognize 
sequences of aligned grains. Some of these 
are track grains; others are spurious grains ({\it fog}) not 
associated to particle tracks. The three-dimensional structure of a track 
in an emulsion layer ({\it microtrack}) is reconstructed by combining 
clusters belonging to images at different levels. Each microtrack 
pair is connected across the plastic base to form the {\it base track}. A set 
of base tracks forms a {\it volume track}.   
 The ESS is based on the use of commercial hardware components. The 
software used for data taking and track reconstruction has a modular 
structure, providing the flexibility needed to upgrade the system 
following the technological progress.

The Japanese S-UTS system, Fig. \ref{fig:ESS} right, is based
on hardware designed and made in Nagoya; the software 
system is mounted in specially designed electronic boards.

\section{Physics performances. Conclusions}
\label{sec:PhysicsPerformance}
\begin{figure}
\begin{center}
\mbox{\epsfig{figure=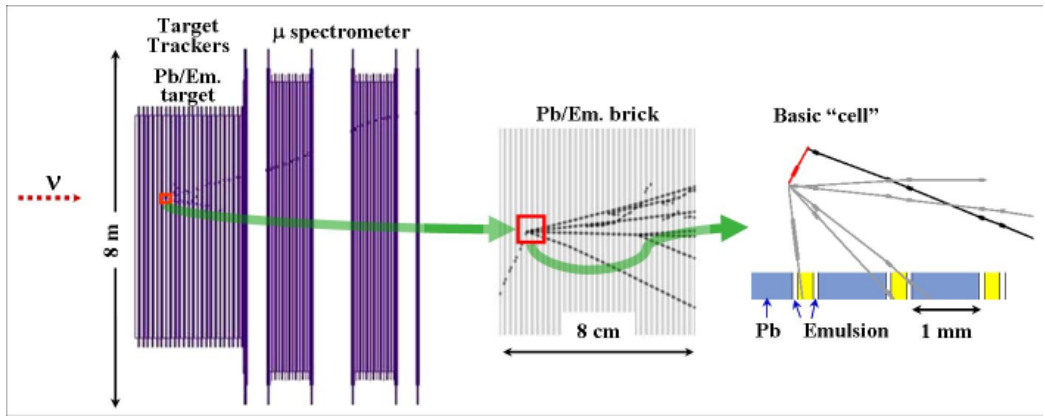,height=5.5cm}}
\caption {Sketch of one supermodule and of one event in the emulsions.}
\label{fig:emul}
\end{center}
\end{figure}

The detection efficiency of tau decays was studied by  
MonteCarlo simulations. One distinguishes two cases: $(i)$ 
``short'' decays: the tau decays in the same lead plate where it is 
produced; the signature is a non-zero impact parameter of the decay 
products with respect to the primary vertex, which can 
 be determined for multi-prong deep inelastic scatterings 
(DIS), Fig. \ref{fig:emul}. $(ii)$ ``long'' decays: the tau   
is measured directly and the kink angle of the charged decays 
accurately determined for DIS and quasi elastic (QE) neutrino 
interactions. 

The first run is scheduled for the second half of 2006. After 
the commissioning of the CNGS beam and of the OPERA electronic detectors 
with a low intensity beam, there will be a normal intensity run.

 The interactions in the bricks allow to check the analysis 
procedure, the vertex finding efficiencies and the beam induced 
background. The expected number of $\tau$ events, for a 
beam intensity of $4.5 \cdot 10^{19}$ pot/year, is 11 (16) events
for $\dm2 = 2.4 \cdot 10^{-3}$ eV$^2$ (for $\dm2 = 3 \cdot 10^{-3}$ eV$^2$).
The background is expected to be $< 1$. It is hoped that one may improve 
the selection by $\sim 30\%$ and that the beam may be increased by a 
factor of 1.5. One should be able to achieve the discovery potential 
in few years. 

We shall also search for $\nmne$ oscillations. In case no $\ne$ is observed 
and assuming $\dm2 = 2.5 \cdot 10^{-3}$ eV$^2$, OPERA shall set 
a limit $\sin^2 2\theta_{13}<0.06$ (90\% C.L.) \cite{t13bib}.

Several byproducts should be obtained with the electronic detectors.

{\normalsize

\section{Acknowledgements}
We would like to acknowledge the cooperation of all the members of the OPERA
Collaboration.

}
\end{document}